\documentclass[usegraphicx,usenatbib,letterpaper]{emulateapj}

\usepackage{ulem}
\usepackage{color}
\usepackage{graphics,graphicx}
\usepackage{times} 
\usepackage{amssymb}
\usepackage{amsmath}
\usepackage{natbib}
\usepackage{url}
\usepackage{multirow}
\usepackage{ctable}
\usepackage{threeparttable}
\newif\ifAMStwofonts
\AMStwofontstrue
\def\xmm{{\it XMM-Newton}}

\def\suzaku{{\it Suzaku}}

\def\epicpn{{EPIC-pn}}
\def\epicmos1{{EPIC-MOS1}}
\def\epicmos2{{EPIC-MOS2}}
\def\epicmos{{EPIC-MOS}}
\def\xis{{\rm XIS}}

\def\nustar{{\it NuSTAR}}
\def\astroh{{\it Astro-H}}

\def\H0{{\rm ~km~s^{-1}~Mpc^{-1}}}

\def\kev{\hbox{\rm keV}}

\def\ergps{\hbox{erg~s$^{-1}$}}

\def\msun{\hbox{$\rm M_{\odot}$}}

\def\xspec{\hbox{\small XSPEC}}

\def\heasoft{\hbox{\rm{\small HEASOFT}}}
\def\nustardas{\rm {\small NUSTARDAS}}

\def\addascaspec{\hbox{\rm{\small ADDASCASPEC~\/}}}
\def\sas{\hbox{\rm{\small SAS~\/}}}
\def\epchain{\hbox{\rm{\small EPCHAIN}}}
\def\emchain{\hbox{\rm{\small EMCHAIN}}}

\def\rmfgen{\hbox{\rm{\small RMFGEN}}}
\def\arfgen{\hbox{\rm{\small ARFGEN}}}

\def\addascaspec{\hbox{\rm{\small ADDASCASPEC}}}
\def\nupipeline{\rm{\small NUPIPELINE}}
\def\nuproducts{\rm{\small NUPRODUCTS}}

\def\grid25{\hbox{\rm{\small GRID25}}}

\def\simpl{\rm{\small SIMPL}}
\def\brems{\rm{\small BREMS}}

\def\tbabs{\rm{\small TBABS}}
\def\tbnew{\rm{\small TBNEW}}
\def\diskbb{\rm{\small DISKBB}}
\def\diskpbb{\rm{\small DISKPBB}}

\def\comptt{\rm{\small COMPTT}}

\def\zsun{$Z_{\odot}$}

\def\tbnewlink{http://pulsar.sternwarte.uni-erlangen.de/wilms/research/tbabs}
\def\suzakuguide{http://heasarc.gsfc.nasa.gov/docs/suzaku/analysis/}
\def\nedlink{http://ned.ipac.caltech.edu/}

\def\eg{{\it e.g.}}

\def\ie{{\it i.e.~\/}}

\def\la{\mathrel{\hbox{\rlap{\hbox{\lower4pt\hbox{$\sim$}}}{\raise2pt\hbox{$<$}}}}}
\def\ga{\mathrel{\hbox{\rlap{\hbox{\lower4pt\hbox{$\sim$}}}{\raise2pt\hbox{$>$}}}}}

\def\d25{D$_{25}$}

\def\.25{0.25 keV\thinspace}

\def\hoii{\rm Holmberg\,II X-1}

\shorttitle{Broadband X-ray observations of \hoii}
\shortauthors{D.~J. Walton et al.}

\begin{document}

\title{\textit{NuSTAR}, \textit{XMM-Newton} and \textit{Suzaku} Observations of the Ultraluminous X-ray Source Holmberg\,II X-1}

\author{D. J. Walton\altaffilmark{1,2},
M. J. Middleton\altaffilmark{3},
V. Rana\altaffilmark{2},
J. M. Miller\altaffilmark{4},
F. A. Harrison\altaffilmark{2},
A. C. Fabian\altaffilmark{3},
M. Bachetti\altaffilmark{5,6},
D. Barret\altaffilmark{5,6},
S. E. Boggs\altaffilmark{7},
F. E. Christensen\altaffilmark{8},
W. W. Craig\altaffilmark{7},
F. Fuerst\altaffilmark{2},
B. W. Grefenstette\altaffilmark{2},
C. J. Hailey\altaffilmark{9},
K. K. Madsen\altaffilmark{2}, \\
D. Stern\altaffilmark{1},
W. Zhang\altaffilmark{10}
}
\affil{
$^{1}$ Jet Propulsion Laboratory, California Institute of Technology, Pasadena, CA 91109, USA \\
$^{2}$ Space Radiation Laboratory, California Institute of Technology, Pasadena, CA 91125, USA \\
$^{3}$ Institute of Astronomy, University of Cambridge, Madingley Road, Cambridge CB3 0HA, UK \\
$^{4}$ Department of Astronomy, University of Michigan, 1085 S. University Ave., Ann Arbor, MI, 49109-1107, USA \\
$^{5}$ Universite de Toulouse; UPS-OMP; IRAP; Toulouse, France \\
$^{6}$ CNRS; IRAP; 9 Av. colonel Roche, BP 44346, F-31028 Toulouse cedex 4, France \\
$^{7}$ Space Sciences Laboratory, University of California, Berkeley, CA 94720, USA \\
$^{8}$ DTU Space, National Space Institute, Technical University of Denmark, Elektrovej 327, DK-2800 Lyngby, Denmark \\
$^{9}$ Columbia Astrophysics Laboratory, Columbia University, New York, NY 10027, USA \\
$^{10}$ NASA Goddard Space Flight Center, Greenbelt, MD 20771, USA \\
}

\begin{abstract}
We present the first broadband 0.3--25.0\,\kev\ X-ray observations of the
bright ultraluminous X-ray source (ULX) \hoii, performed by \nustar, \xmm\
and \suzaku\ in September 2013. The \nustar\ data provide the first
observations of \hoii\ above 10\,\kev, and reveal a very steep high-energy
spectrum, similar to other ULXs observed by \nustar\ to date. These
observations further demonstrate that ULXs exhibit spectral states that are not
typically seen in Galactic black hole binaries. Comparison with other sources
implies that \hoii\ accretes at a high fraction of its Eddington accretion
rate, and possibly exceeds it. The soft X-ray spectrum ($E<10$\,\kev) appears to
be dominated by two blackbody-like emission components, the hotter of which 
may be associated with an accretion disk. However, all simple disk models
under-predict the \nustar\ data above $\sim$10\,\kev\ and require an additional
emission component at the highest energies probed, implying the \nustar\ data
does not fall away with a Wien spectrum. We investigate physical
origins for such an additional high-energy emission component, and favor a
scenario in which the excess arises from Compton scattering in a hot corona of
electrons with some properties similar to the very-high state seen in Galactic
binaries. The observed broadband 0.3--25.0\,\kev\ luminosity inferred from these
epochs is $L_{\rm{X}}=(8.1\pm0.1)\times10^{39}$\,\ergps, typical for \hoii, with the
majority of this flux ($\sim$90\%) emitted below 10\,\kev.
\end{abstract}


\section{Introduction}

Ultraluminous X-ray Sources (ULXs) are off-nuclear point sources in external
galaxies that radiate in excess of the Eddington limit for a $\sim$10\,\msun\
`stellar' black hole such as those found in Galactic black hole binary systems
(BHBs; \eg\ \citealt{Orosz03}), \ie $L_{\rm{X}}>10^{39}$\,\ergps\
(\citealt{Roberts07rev, Feng11rev}). These luminosities may result from either
the presence of larger black holes than those observed in our own Galaxy (\eg\
\citealt{Miller04, Zampieri09}), or from super-Eddington modes of accretion
(\eg\ \citealt{Poutanen07}). The majority of ULXs only radiate marginally in
excess of $10^{39}$\,\ergps\ (\citealt{WaltonULXcat, Swartz11}), and there is
now a reasonable consensus that these sources likely represent a high
luminosity extension of the stellar remnant BHB population 
(\citealt{Middleton13nat, Liu13nat, Motch14nat}). The best candidates for black
holes significantly more massive than Galactic BHBs are instead found among
the brightest members of the ULX population, with luminosities
$L_{\rm{X}}>10^{40}$\,\ergps\ (\eg\ \citealt{Farrell09, Sutton12,
Pasham14nat}). However, while the majority of origins considered for this high
luminosity population have focused on accretion onto black holes, with good
reason given the luminosities in question, we now know that one source that
exhibits these luminosities (albeit transiently) is in fact a pulsar
(\citealt{Bachetti14nat}), further expanding the pool of plausible scenarios (see
also \citealt{King01}). 

\hoii\ is a source of particular interest among the bright ($L_{\rm{X}} >
10^{40}$\,\ergps) ULX population. Although variable on moderate timescales
($\sim$days--weeks, and longer; \eg\ \citealt{Grise10}), it persistently radiates
at a luminosity of $L_{\rm{X}}\sim10^{40}$ \ergps, and is known to
display a cool thermal component which may evolve following the $L \propto
{T}^{4}$ relation expected for stable blackbody emission (\citealt{Feng09,
Miller13ulx}). If associated with a standard accretion disc that extends close to
the black hole, this would imply the presence of a massive black hole. However,
high signal-to-noise X-ray observations of \hoii\ show the presence of curvature
in the $\sim$3--10\,\kev\ continuum, which is generally interpreted as a
signature of high/super-Eddington accretion (\citealt{Gladstone09, Kajava12};
see also \citealt{Motch14nat}), although definitive evidence that this represents
a genuine spectral cutoff is currently lacking for \hoii\ owing to the limited
bandpass previously available (\citealt{Caball10, Walton4517}).

Beyond the X-ray regime, observations of \hoii\ at longer wavelengths have
revealed an X-ray ionised nebula, and emission from moderately ionised oxygen
which, through photon counting arguments, confirm the extreme X-ray
luminosity and rule out strongly beamed emission, implying that X-ray flux is
broadly isotropic (\citealt{Berghea10b}). In addition, radio observations have
now revealed that \hoii\ repeatedly launches collimated jets (\citealt{Cseh14}),
making it the only confirmed ULX known to launch such outflows.

Here, we present the results from broadband X-ray observations of \hoii\
undertaken with the \nustar\ (\citealt{NUSTAR}), \xmm\ (\citealt{XMM}) and
\suzaku\ (\citealt{SUZAKU}) observatories. The paper is structured as follows: in
section \ref{sec_red} we describe our observations and data reduction, in section
\ref{sec_spec} we present our analysis of these data, and in section \ref{sec_dis}
we discuss our results and draw conclusions.

\section{Observations and Data Reduction}
\label{sec_red}

During September 2013 \nustar\ and \xmm\ performed two coordinated
observations of \hoii, separated by $\sim$8 days. The \xmm\ exposures were
short owing to visibility limitations, but both were simultaneous with some
portion of the longer \nustar\ observations, providing soft X-ray coverage down
to $\sim$0.3\,\kev. \suzaku\ also performed two observations of \hoii\ in
September 2013, the first coordinated with the second \nustar+\xmm\
observation, and the second $\sim$10 days later. The details of the observations
performed in our 2013 campaign are given in Table \ref{tab_obs} (although the
first \nustar\ epoch is comprised of two OBSIDs, it is one continuous
observation). For each mission, source products were extracted from circular
regions (\nustar: radius $\sim$75$''$; \xmm: $\sim$40$''$ for \epicpn,
$\sim$50$''$ for \epicmos; \suzaku: $\sim$230$''$), and the backgrounds were
estimated from blank regions free of contaminating sources on the same
detector as \hoii.

\subsection{NuSTAR}

We reduced the \nustar\ data using the standard pipeline, \nupipeline, part of the
\nustar\ Data Analysis Software (\nustardas, v1.3.1; included in the \heasoft\
distribution), with the \nustar\ instrumental calibration files from caldb
v20140414. The unfiltered event files were cleaned with the standard depth
correction, significantly reducing the internal high-energy background, and
passages through the South Atlantic Anomaly were removed. Source spectra and
instrumental responses were produced for each of the two focal plane modules
(FPMA/B) using \nuproducts. The FPMA and FPMB data each provide an
independent detection up to $\sim$25\,\kev.

\subsection{XMM-Newton}

The \xmm\ data reduction was carried out with the \xmm\ Science Analysis
System (\sas v13.5.0), following the standard prescription provided in the online
guide.\footnote{http://xmm.esac.esa.int/} Raw data files were cleaned using
\epchain\ and \emchain\ for the \epicpn\ and \epicmos\ detectors, respectively.
Only single and double events were considered for \epicpn\ and single to
quadruple events for \epicmos. Periods of high background flares were excluded.
Instrumental response files were generated with \rmfgen\ and \arfgen. After
performing the reduction separately for the two \epicmos\ detectors, and
confirming their consistency, these data were combined into a single spectrum
using \addascaspec. The \xmm\ data are analysed over the 0.3--10.0\,\kev\
energy range.

%\pagebreak
\subsection{Suzaku}

As we have high energy coverage from \nustar, we only use the data obtained
with the XIS detectors, and do not consider the non-imaging HXD PIN detector.
We cleaned the unfiltered event files with the latest calibration and the standard
screening criteria for each of the \xis\ CCDs (XIS0/1/3) and editing modes
operated (3x3/5x5), using the latest \heasoft\ package (v6.15) as recommended
in the \suzaku\ Data Reduction Guide.\footnote{\suzakuguide} Instrumental
responses were generated for each detector using {\small XISRESP} (with a
medium resolution). Finally, after checking their consistency, the data from the
front-illuminated detectors (XIS0/3) were combined using \addascaspec. The
front- and back-illuminated (FI and BI) data are analysed over the 0.7--10.0 and
0.7--8.0\,\kev\ energy ranges, respectively.

\begin{table}
  \caption{Details of the 2013 X-ray observations of Holmberg\,II X-1 considered
  in this work.}
\begin{center}
\begin{tabular}{c c c}
\hline
\hline
\\[-0.25cm]
OBSID & Start Date & Good Exposure\tmark[a] \\
& & (ks) \\
\\[-0.3cm]
\hline
\hline
\\[-0.1cm]
\multicolumn{3}{c}{\textit{NuSTAR}} \\
\\[-0.2cm]
30001031002/3 & 2013-09-09 & 111 \\
\\[-0.225cm]
30001031005 & 2013-09-17 & 111 \\
\\[-0.1cm]
\multicolumn{3}{c}{\textit{XMM-Newton}} \\
\\[-0.2cm]
0724810101 & 2013-09-09 & 5/8 \\
\\[-0.225cm]
0724810301 & 2013-09-17 & 6/9 \\
\\[-0.1cm]
\multicolumn{3}{c}{\textit{Suzaku}} \\
\\[-0.2cm]
708015010 & 2013-09-17 & 52 \\
\\[-0.225cm]
708015020 & 2013-09-27 & 49 \\
\\[-0.2cm]
\hline
\hline
\\[-0.15cm]
\end{tabular}
\\
$^{a}$ \xmm\ exposures are listed for the \epicpn/MOS detectors.
\vspace*{0.3cm}
\label{tab_obs}
\end{center}
\end{table}

\section{Initial Spectral Analysis}
\label{sec_spec}

\begin{figure*}
\hspace*{-0.5cm}
\epsscale{0.56}
\plotone{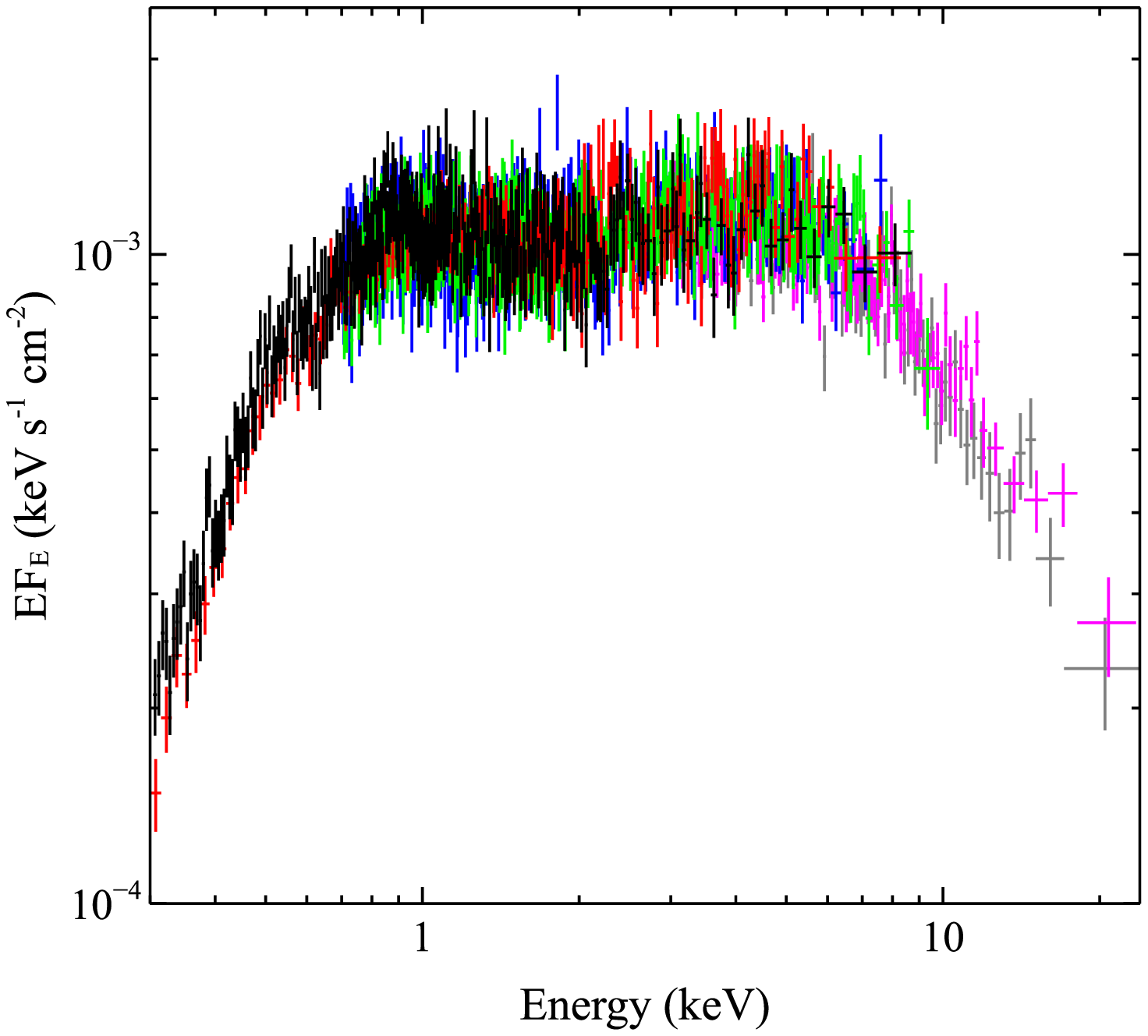}
\hspace*{0.5cm}
\plotone{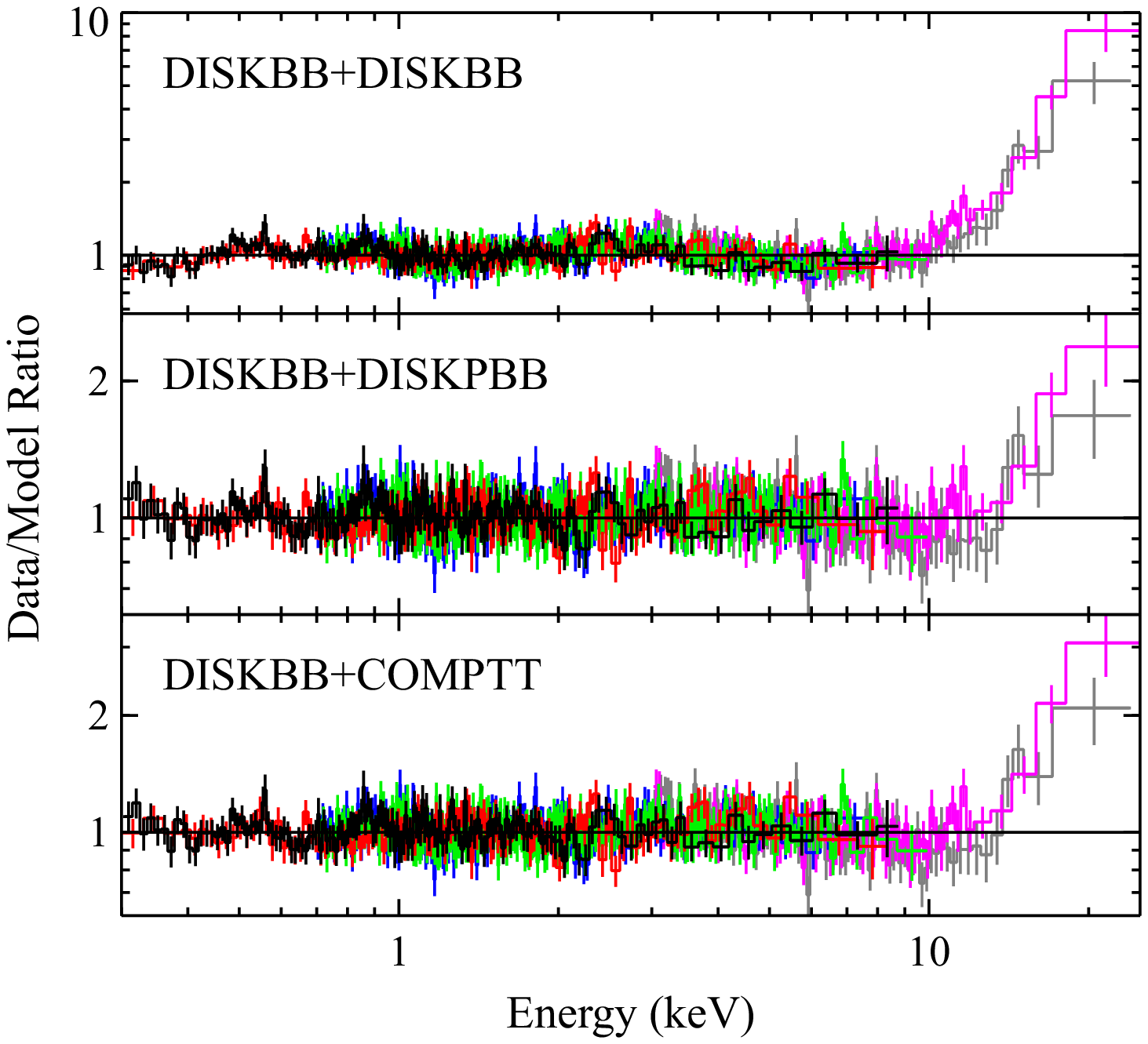}
\caption{
\textit{Left panel:} the broadband X-ray spectrum of \hoii, observed by \nustar\
(FPMA in magenta, FPMB in grey), \xmm\ (\epicpn\ in black, \epicmos\ in red)
and \suzaku\ (FI XIS in green, BI XIS in blue) in 2013, unfolded through a model
simply consisting of a constant. The \nustar\ data clearly demonstrate the
presence of a high-energy spectral cutoff. \textit{Right panel:} data/model ratios
for the three two-component thermal continuum models considered. Each results
in a clear excess in the residuals at high energies.}
\label{fig_spec}
\end{figure*}

The spectra obtained from each of the two \nustar+\xmm\ observations are
consistent with each other, as are the spectra from each of the two \suzaku\
observations; no significant spectral variability is observed among any of
these epochs. Therefore, to simplify the analysis, we combine the data into
time-averaged spectra, and model the \nustar, \xmm\ and \suzaku\ data
simultaneously; the spectral agreement between \nustar, \xmm\ and
\suzaku\ in their common 3--10\,\kev\ bandpass is known to be good to
within $\sim$10\% in absolute normalisation and a few \% in spectral
agreement (within errors, \citealt{NUSTARcal}; see also \citealt{Walton13culx,
Walton14hoIX, Brenneman14b}). All spectral analysis is performed with
\xspec\ v12.8.1 (\citealt{xspec}). The \xmm\ and \suzaku\ datasets are
rebinned to 50 counts per bin, while the \nustar\ data are rebinned to 100
counts per bin, owing to the larger relative contribution of the instrumental
background at higher energies. The observed broadband X-ray spectrum is
shown in Figure \ref{fig_spec}.

Some basic conclusions can immediately be drawn from visual inspection of this
spectrum. The continuum above $\sim$2\,\kev\ is not powerlaw-like. The
\nustar\ data clearly confirm the presence of the curvature in the
$\sim$3--10\,\kev\ bandpass claimed previously (\eg\ \citealt{Stobbart06,
Gladstone09}), and that this is a genuine spectral cutoff. This is broadly similar
to other ULXs observed by \nustar\ to date (\citealt{Bachetti13, Rana15,
Walton13culx, Walton14hoIX, Walton15}). Our analysis of these other sources has
generally found that two blackbody-like thermal components are required to
model their broadband spectra (see also \citealt{Miller14}), and visual evidence
for distinct spectral components above and below $\sim$2\,\kev\ can also be
seen here. We therefore focus on similar models in our analysis of \hoii.

All subsequent models include neutral absorption from both the Galactic column
towards Holmberg\,II ($N_{\rm{H;Gal}}=3.7\times10^{20}$ cm$^{-2}$;
\citealt{NH}), and an intrinsic absorption column at the redshift of Holmberg\,II
($z=0.002225$)\footnote{from the NASA Extragalactic Database: \nedlink}
which is free to vary. Both are modeled with \tbnew\
(\citealt{tbabs})\footnote{\tbnewlink}. Unless stated otherwise, these absorption
components are assumed to have solar abundances, and we adopt the abundance
set in \cite{tbabs} and the cross-sections of \cite{Verner96}. Parameter uncertainties
are quoted at 90\% confidence for one parameter of interest throughout, and we
account for residual cross-calibration flux uncertainties between different detectors
by allowing multiplicative constants to float between them, fixing \epicpn\ to unity.

We begin by confirming quantitatively that two blackbody-like continuum
components are required. Simple accretion disk models (\eg\ \diskbb,
\citealt{diskbb}; \diskpbb, \citealt{diskpbb}) cannot fit the broadband data by
themselves, as they cannot reproduce the double-peaked spectral shape
below 10\,\kev. We therefore apply models that include two continuum
components, starting with a model consisting of two \diskbb\ components.
However, the fit is poor, with $\chi^{2}$/degrees of freedom (DoF) = 2401/1899,
and this model leaves a strong excess in the \nustar\ data above $\sim$10\,\kev\
(see Figure \ref{fig_spec}). Although the fit is improved, this excess is not fully
resolved by allowing the radial temperature profile of the hotter of the two disk
components to vary (using \diskpbb; $\chi^{2}$/DoF = 1938/1898). Nor is it fully
resolved by replacing the hotter disk component with a thermal Comptonization
model (\comptt; \citealt{comptt}), although the fit is again improved over the two
\diskbb\ model ($\chi^{2}$/DoF = 1979/1898). For the Comptonization model,
we link the seed photon temperature for the Comptonization to that of the lower
temperature \diskbb\ component, as would be appropriate for standard,
optically thin Comptonization. However, we find the corona to be cool and
optically thick ($\tau\sim6$; $T_{\rm{in}}\sim2$\,\kev), similar to the results
typically found for ULXs (\eg\ \citealt{Stobbart06, Walton4517}). These
parameters result in a disk-like spectrum with a thermal roll-over above
$\sim$5\,\kev. If the corona is optically thick and shrouds the inner disk, our
assumption regarding the seed photon temperature may not be appropriate
(\eg\ \citealt{Gladstone09}). However, even if we unlink this temperature from
that of the \diskbb\ component, the hard excess still persists.

\section{The Hard Excess}
\label{sec_hardex}

Evidence for similar hard excesses has been seen in other \nustar\ observations
of ULXs (\eg\ \citealt{Walton13culx, Walton14hoIX, Rana15, Mukherjee15}).
However, for the cases that require complex thermal continua below
$\sim$10\,\kev\ (\ie\ Holmberg\,IX X-1, IC\,342 X-1 and NGC\,5204 X-1) the
significance of the excess is model dependent. Here, we find that a complex
continuum consisting of two blackbody-like thermal components is required to
model the spectrum below $\sim$10\,\kev, and that a hard excess remains in
the \nustar\ data above $\sim$10\,\kev\ regardless of the details of this lower
energy continuum. An additional high-energy continuum component is required
by any model that falls away with a thermal Wien tail. To address the nature of
this hard excess, we focus on the best fitting two-component thermal model
discussed above, the \diskbb+\diskpbb\ combination.

\begin{figure*}
\hspace*{-0.5cm}
\epsscale{0.56}
\plotone{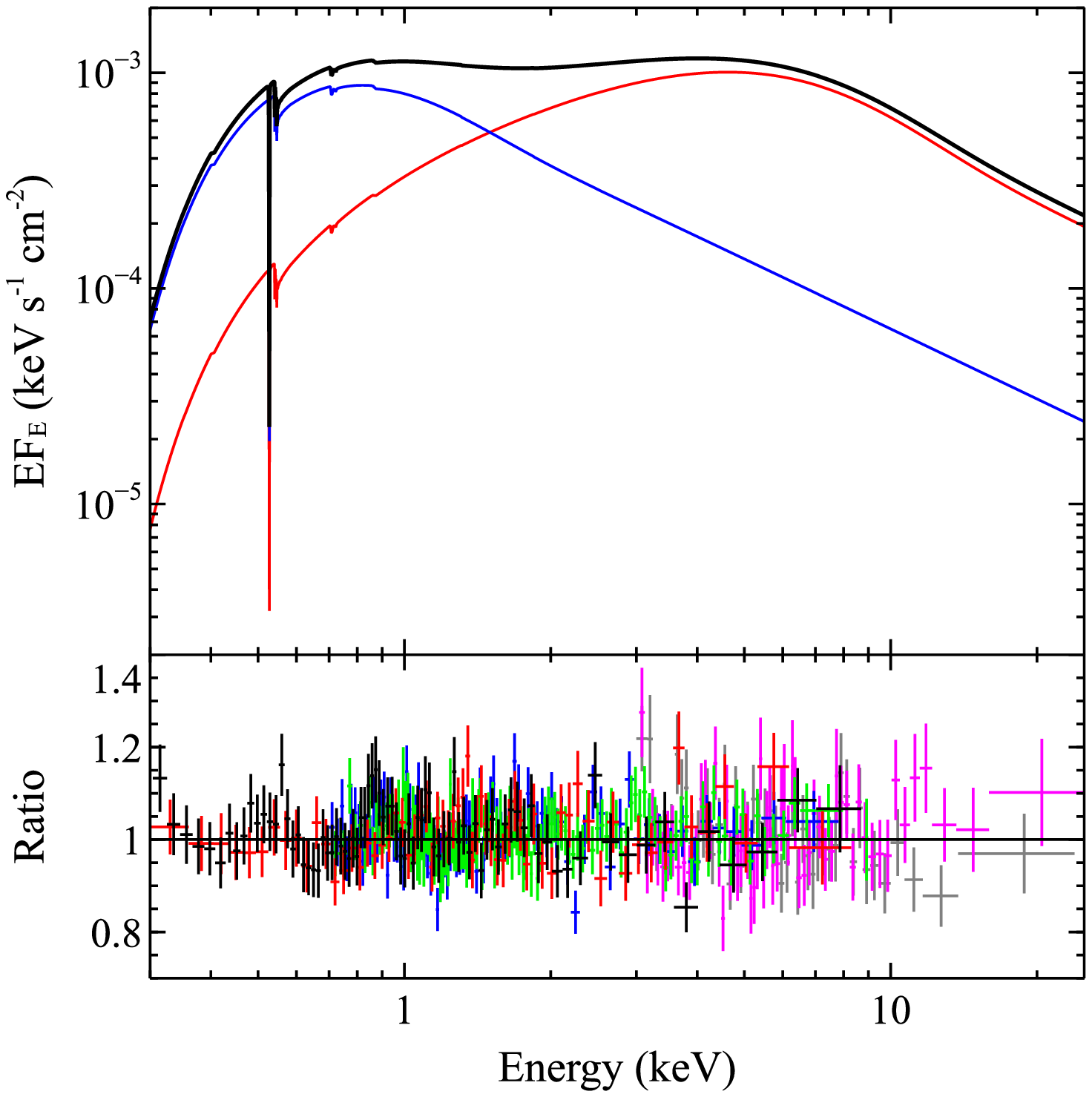}
\hspace*{0.5cm}
\plotone{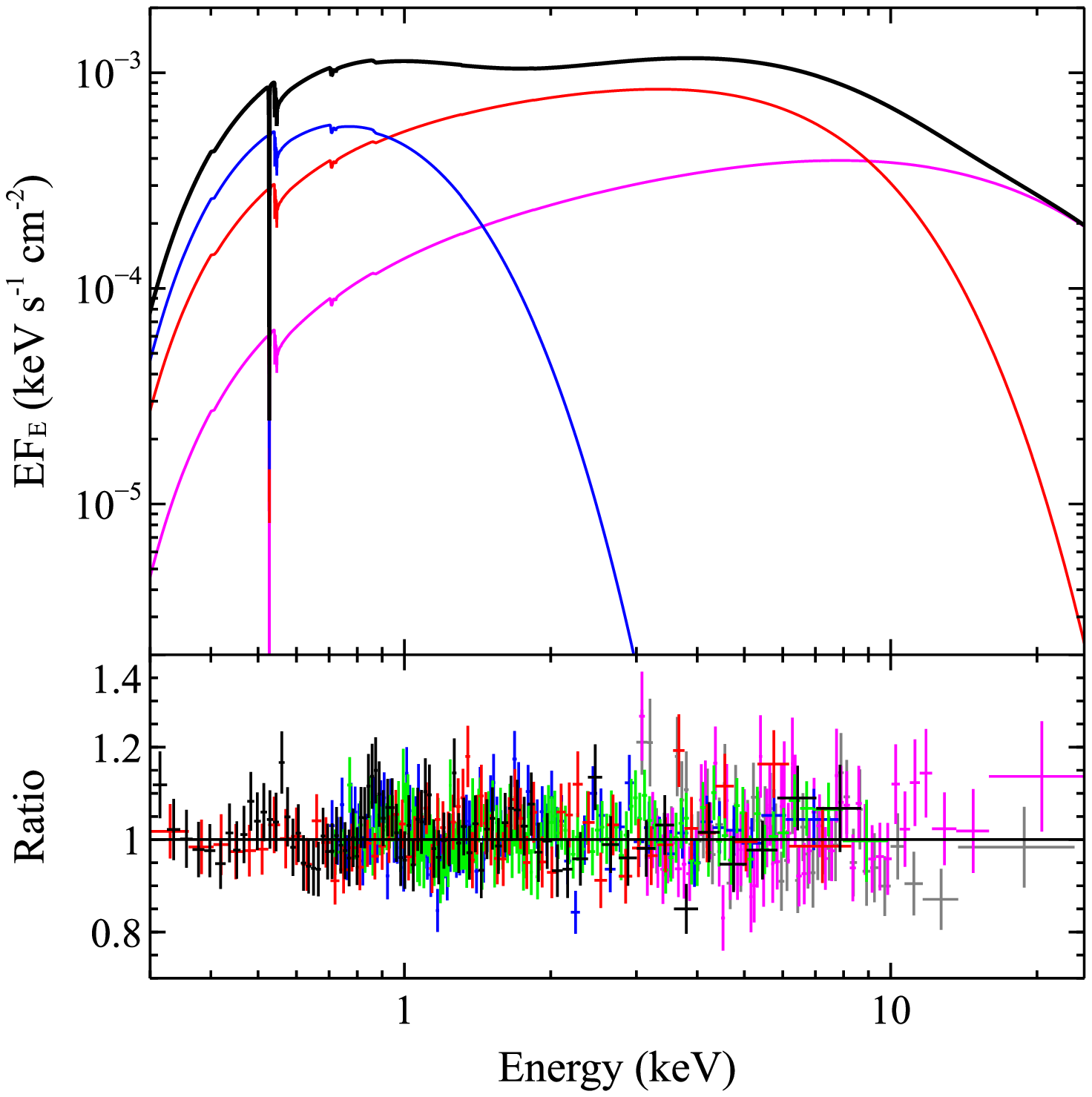}
\caption{
Fits to the broadband spectrum of \hoii\ with our three-component continuum
models. The top panels show the models applied (black line); the base continuum
model is a combination of DISKBB (blue) and DISKPBB (red) disk components, and
the left panels show the hard excess modeled with a powerlaw tail (SIMPL), while
the right panels show this excess modeled with bremsstrahlung emission
(magenta). The bottom panels show their data/model ratios, with the same color
scheme as Figure \ref{fig_spec}.
}
\label{fig_models}
\end{figure*}

To account for this hard excess, we first consider whether it might arise from
an optically thin, hot Comptonizing corona such as found in Galactic BHBs. We
add an additional high-energy powerlaw continuum using \simpl\
(\citealt{simpl}). This is a convolution model that scatters some fraction
($f_{\rm{scat}}$) of an input spectrum into a high-energy powerlaw tail with
photon index $\Gamma$. We initially apply \simpl\ to both disk components,
such that each contributes the same powerlaw tail (i.e. $\Gamma$ and
$f_{\rm{scat}}$ are the same for each component). This would correspond to
the `patchy disk' scenario outlined in \cite{Miller14} (see discussion), in which
the hotter and cooler emission regions are co-located, and thus the scattering
medium likely subtends a similar solid angle for each. This gives an excellent
fit to the broadband spectrum with $\chi^{2}/\rm{DoF} = 1884/1896$,
providing a statistical improvement of $\Delta\chi^{2}=54$ for two additional
free parameters. The broadband 0.3--25.0\,\kev\ flux obtained with this
model corresponds to an observed luminosity of $L_{\rm{X}} = (8.1 \pm 0.1)
\times 10^{39}$\,\ergps\ for a distance of 3.39\,Mpc
(\citealt{Karachentsev02}). Correcting for the neutral absorbing column
inferred, the implied intrinsic 0.3--25.0\,\kev\ luminosity is $L_{\rm{X,int}} =
(10.1 \pm 0.4) \times 10^{39}$\,\ergps. Best fit model parameters are given
in Table \ref{tab_param}, and the fit is shown in Figure \ref{fig_models}.

The \simpl\ parameters are poorly constrained owing to the limited high-energy
statistics, resulting in strong degeneracies. However, the best-fit photon index is
steep, $\Gamma=3.1^{+0.3}_{-1.2}$, similar to the coronae observed from
Galactic binaries in the very-high state (\citealt{Remillard06rev}). With the addition
of this powerlaw continuum, we find that the radial temperature index obtained
for the hotter (\diskpbb) component, $p=0.67^{+0.10}_{-0.05}$, is consistent
with that expected for a standard thin accretion disk ($p=0.75$;
\citealt{Shakura73}). If we fix $p=0.75$, the constraint on the photon index is
significantly stronger: $\Gamma=3.2^{+0.2}_{-0.1}$. We note that an equally
good fit can be obtained assuming only to the (hotter) \diskpbb\ component is
scattered into a high-energy tail. This would correspond to an alternative scenario
in which the hotter and cooler emission regions are not co-spatial, but the cooler
emission comes from much larger radii, and thus the scattering region, which is
likely centrally located, does not intercept a significant fraction of this emission.
The \simpl\ parameters obtained are similarly poorly constrained and remain
consistent with those presented, but in this case we find $p=0.55\pm0.01$,
significantly shallower than the thin disk case, closer instead to what might be
expected from an accretion disk experiencing significant photon advection (\eg\
\citealt{Abram88}).

\begin{table}
  \caption{Key parameters obtained for the three-component continuum models
 for Holmberg II X-1}
\begin{center}
\begin{tabular}{c c c c}
\hline
\hline
\\[-0.15cm]
Model & Parameter & \multicolumn{2}{c}{Hard continuum} \\
Component & & Powerlaw & Bremsstrahlung \\
\\[-0.2cm]
\hline
\hline
\\[-0.1cm]
\tbabs\ & $N_{\rm{H}}$ [$10^{20}~\rm{cm}^{-2}$] & $6^{+2}_{-1}$ & $6^{+2}_{-1}$ \\
\\[-0.225cm]
\diskbb\ & $T_{\rm{in}}$ [keV] & $0.20^{+0.03}_{-0.04}$ & $0.23^{+0.01}_{-0.02}$ \\
\\[-0.225cm]
\diskpbb\ & $T_{\rm{in}}$ [keV] & $1.8^{+0.7}_{-0.3}$ & $2.0^{+0.3}_{-0.2}$ \\
\\[-0.225cm]
& $p$ & $0.67^{+0.10}_{-0.05}$ & $0.55^{+0.03}_{-0.01}$ \\
\\[-0.225cm]
\simpl\ & $\Gamma$ & $3.1^{+0.3}_{-1.2}$ & -- \\
\\[-0.225cm]
& $f_{\rm{scat}}$ [\%] & $40^{+50}_{-30}$ & -- \\
\\[-0.225cm]
\brems\ & $T_{\rm{brems}}$ [keV] & -- & $12^{+32}_{-3}$ \\
\\[-0.2cm]
\hline
\\[-0.15cm]
$\chi^{2}$/DoF & & 1884/1896 & 1884/1896 \\
\\[-0.2cm]
\hline
\hline
\\[-0.15cm]
\end{tabular}
\label{tab_param}
\end{center}
\end{table}

An alternative possibility for the hard excess is that it is associated with the
radio jets rather than with a Comptonizing corona. Several authors have
suggested that SS433 might be a Galactic ULX analog, merely observed at a
sufficiently high inclination that the X-ray emission from the central regions
of the accretion flow is obscured from our view (\citealt{Begelman06}). This
source is also known to launch collimated jets, which themselves emit X-rays
with a spectrum comprised of strong line emission (iron K$\alpha$ equivalent
width of $\sim$350\,eV) on top of a bremsstrahlung continuum (\eg\
\citealt{Marshall02, Lopez06}). We therefore fit a model including a
bremsstrahlung continuum for the hard excess to test an origin from
SS433-like jet emission.

While the limits for any narrow iron features from these data are not as
stringent as for other ULXs (\eg\ \citealt{Walton13hoIXfeK}), they are sufficient
to exclude emission lines of the strength observed in SS433, with important
implications for this jet emission model. We find that any narrow emission lines
in the immediate 6--7\,\kev\ iron K$\alpha$ bandpass must have equivalent
widths less than 40\,eV (at 90\% confidence).

The metallicity of the system is a key issue relevant to the absence of strong jet
lines in \hoii. The dwarf galaxy Holmberg\,II appears to have a sub-solar
metallicity in general, with an estimated metalicity of $\sim$0.2 $Z_{\rm{\odot}}$
(\citealt{Egorov13}). A low metallicity could help hide any line emission.
However, the low average metallicity of Holmberg\,II is not necessarily
representative of the \hoii\ system, which may, for example, have been locally
enriched by the supernova event in which the central compact object was
formed. Should the excess absorption over the Galactic column along our
line-of-sight to \hoii\ be largely local to the system, this would provide a
better estimate of the relevant metallicity than looking at the galaxy as a whole.
Unfortunately the lowest energies ($E<0.7$\,\kev) of the broadband spectra
obtained in our campaign, which are the most sensitive to the details of the
absorption model, are only covered by the \xmm\ observations, which were
very short. These data are not able to provide any strong constraints on
elemental abundances for the absorption intrinsic to Holmberg\,II, either
individually or collectively; even assuming a consistent abundance relative to
solar for all elements heaver than carbon we only find a weak upper limit of
$Z<1.8$\,\zsun, although we note that the best-fit obtained with this
approach implies a solar metallicity.

In their analysis of the longest \xmm\ observation of \hoii\ available to date,
\cite{Goad06} suggest that the oxygen abundance of the absorbing medium
toward \hoii\ might be sub-solar (O/solar $\sim$ 0.6), based on the neutral
oxygen edge at $\sim$0.55\,\kev. However, this is estimated with a steep
($\Gamma = 2.6$) powerlaw continuum extrapolated to arbitrarily low
energies, and our broadband observations show that the continuum is not a
powerlaw. This extrapolation likely results in the absorption column being
overestimated, which would in turn lead to the oxygen abundance being
underestimated in order to match the depth of the edge. Indeed, when these
authors consider models in which the low-energy continuum is a disk
blackbody, a solar abundance gives an improved fit over the sub-solar
abundance inferred with the powerlaw continuum. \cite{Winter07} also
analyzed this dataset with a view to constraining the abundances of oxygen
and iron, treating the soft emission both as a blackbody and a disk blackbody,
and typically found solar or slightly super-solar oxygen abundances,
depending on the exact model used (iron abundances were generally poorly
constrained).

In contrast, \cite{Marshall13} find that the outflowing material in the SS433
jets has a strongly super-solar metallicity of $\sim$6 $Z_{\rm{\odot}}$.
Although the metalliticy of the \hoii\ system is ultimately still an open
question, we therefore assume that for SS433-like jet emission, the iron
lines would be a factor of $\sim$5--6 weaker relative to the bremsstrahlung
continuum in \hoii\ than in SS433. However, given the limits discussed above
and the lines observed from SS433, this difference by itself would not be
sufficient to reduce the expected strength of these lines to the point that they
would not be detectable with these observations of \hoii.

We therefore conclude that the jet emission cannot dominate the continuum
emission across the iron bandpass. When applying the bremsstrahlung model
we therefore require that the bremsstrahlung flux is less than that of the hotter
disk component at these energies. SS433-like jet lines would then be
sufficiently diluted by this additional continuum emission to remain undetected,
assuming the difference in metallicity discussed above. This provides an equally
good fit to the powerlaw continuum, with $\chi^{2}/\rm{DoF}=1884/1896$;
best-fit parameters for this model are also given in Table \ref{tab_param}, and
the fit is also shown in Figure \ref{fig_models}. Similar to the coronal parameters
in the previous model, the bremsstrahlung temperature is poorly constrained
owing to remaining parameter degeneracies: $T_{\rm{brems}} =
12^{+32}_{-3}$\,keV. In this case, we find $p=0.55^{+0.03}_{-0.01}$ for the
(hotter) \diskpbb\ component, again significantly shallower than the thin disk
case.

\section{Discussion and Conclusions}
\label{sec_dis}

We have presented a broadband spectral analysis of the ULX \hoii, observed with
\nustar, \xmm\ and \suzaku\ in 2013, which provides the first constraints on
the hard ($E>10$\,\kev) X-ray emission from this source. These observations
were taken at three epochs spanning a period of $\sim$3 weeks. We find no
evidence for significant variability between them, and we therefore focus our
analysis on the average spectrum. As with other ULXs observed with \nustar, we
find clear evidence for a spectral cutoff above $\sim$5\,\kev\ (Figure
\ref{fig_spec}; see \citealt{Bachetti13, Rana15, Walton13culx, Walton14hoIX,
Walton15}), confirming the previous indication from archival \xmm\ observations
(e.g. \citealt{Stobbart06, Gladstone09, Kajava12}). For comparison with previous
observations, we note that $\sim$90\% of the observed broadband luminosity
($L_{\rm{X}} = (8.1 \pm 0.1)\times10^{39}$\,\ergps) is emitted below 10\,\kev,
consistent with the typical range of fluxes for this source (\citealt{Grise10}).

The observed broadband spectrum is not consistent with the standard
accretion states observed at low Eddington rates in Galactic binaries
(\citealt{Remillard06rev}). Under the assumption that these accretion regimes
are mass-independent, this would imply that \hoii\ is accreting at a high fraction
of its Eddington rate, and possibly exceeding it. Although this evidence is
indirect, we note that the behaviour of the best IMBH candidate to date,
ESO\,243-49 HLX1, suggests this is a reasonable assumption to make
(\citealt{Servillat11}). In additon, the spectrum of \hoii\ bears some broad
similarity to the spectrum of the less luminous ULX P13 in NGC\,7793
($L_{\rm{X,peak}} \sim 5 \times 10^{39}$\,\ergps). This source is notable, as
the current dynamical mass constraints for P13 imply a mass of
$\lesssim$15\,\msun\ for the accretor (\citealt{Motch14nat}), indicating it is
accreting at a high/super-Eddington rate. \hoii\ is typically observed to be more
luminous, which may imply a slightly larger Eddington ratio ($L/L_{\rm{E}}$) or
a slightly elevated black hole mass in comparison to P13. Most models for
strong super-Eddington accretion invoke thick accretion disks, resulting in some
level of anisotropic emission, with larger scale heights and therefore increasing
levels of anisotropy at higher accretion rates (\eg\ \citealt{King01, Dotan11}). In
principle, the roughly isotropic emission implied by the ionised nebula
surrounding \hoii\ may therefore imply some upper limit to $L/L_{\rm{E}}$.
However, the uncertain solid angle subtended by the nebula, which appears to
exhibit some mild anisotropy (\eg\ \citealt{Kaaret04}), likely still allows for some
moderate geometrical beaming, so unfortunately it is difficult to be quantitative
here.

The spectrum below 10\,\kev\ shows evidence for two blackbody like thermal
continuum components. However, the Wien tail in all such models falls away
faster than the high-energy data, resulting in a clear excess in the residuals
above $\sim$10\,\kev. These models therefore require an additional,
high-energy spectral component. Similar excesses have
been seen in other ULXs observed by \nustar\ (\citealt{Walton13culx}), but for
the cases where the lower energy spectrum is complex the requirement for an
additional component was model dependent (\eg\ \citealt{Walton14hoIX,
Rana15, Mukherjee15}). That is not the case here, despite the requirement for
a two-component continuum model for the soft X-ray spectrum. This excess
can be modeled successfully with an additional steep powerlaw tail, consistent
with an optically thin Comptonizing corona. It can also formally be modeled
with a hot bremsstrahlung continuum potentially associated with emission
from the radio jets (\citealt{Cseh14}), analogous to that seen from the jets in
SS433 (\citealt{Marshall02, Marshall13}).

In the jet emission model, we find the 2--10\,\kev\ luminosity associated with
the bremsstrahlung component to be $L_{\rm{brems}} = (1.4\pm1.0) \times
10^{39}$\,\ergps\ (the upper limit being a result of our requirement that this
component be fainter than the \diskpbb\ component in the immediate iron
bandpass; see Section \ref{sec_hardex}), $\sim$3 orders of magnitude or more
in excess of the jet emission observed from SS433 ($L_{\rm{brems}}\sim 3
\times 10^{35}$\,\ergps; \citealt{Marshall02}). The bremsstrahlung emission
may scale with the radiative power of the jet. However, the core radio luminosity
($L_{\rm{R}}$) is $\sim$2 orders of magnitude greater in \hoii\ than in SS433
(\citealt{MillerJones08, Cseh14}), implying that $L_{\rm{brems}}/L_{\rm{R}}$
is at least an order of magnitude greater in \hoii\ than in SS433. While there
is the obvious caveat that we are not comparing simultaneous radio and X-ray
observations, the observed X-ray flux during this epoch is typical of this source
(\citealt{Grise10}), and while the variability properties of the radio core are not
currently known, the triple-radio structure implies repeated jet emission
(\citealt{Cseh14}), broadly similar to SS433. However, as the X-ray emission
from SS433 is thought to arise through plasma collisions within the jet, it may
be more likely that the bremsstrahlung emission would scale with the kinetic
power, $L_{\rm{kin}}$, of the jet instead, and $L_{\rm{brems}}/L_{\rm{kin}}$
may therefore be a more appropriate quantity to compare. Unfortunately neither
SS433 or \hoii\ have robust constraints for $L_{\rm{kin}}$ at the time of writing
(\citealt{Marshall13, Cseh14}), so we are not currently able to meaningfully
assess whether these systems are similar in this regard.

Although it is therefore difficult to exclude the SS433-like jet interpretation, high
accretion rate Galactic BHBs in the very-high state exhibit steep ($\Gamma \sim
2.5$) high-energy powerlaw emission that can extend up to $\sim$MeV energies
(\eg\ \citealt{Tomsick99, Remillard06rev}). We therefore favour a similar
interpretation for \hoii, in which the high-energy emission is a powerlaw tail to
the thermal continuum. The smooth nature of the high-energy spectrum and the
transition from the hotter blackbody component to the hard excess probably also
supports this interpretation over one invoking two essentially unrelated emission
components. The likely origin of this powerlaw tail is Comptonization by a hot (or
even non-thermal) coronal plasma. This implies that even though the
3--10\,\kev\ emission can formally be modeled with an optically thick \comptt\
component, it is not physically associated with the Comptonizing corona.
Therefore the thermal continuum likely arises from a high-Eddington accretion
disk (\eg, \citealt{Poutanen07, Middleton11a, Miller14}).

The physical nature of the soft blackbody component remains uncertain. Given the
lack of strong variability between our observations we cannot differentiate between
interpretations invoking disk emission, \eg\ the `patchy disk' scenario recently
proposed by \citealt{Miller14} in which the surface of the accretion disk is
inhomogeneous, with a variety of hot patches embedded across a range of radii
within a cooler surrounding medium, resulting in a more complex temperature
profile than predicted by simple disk models) or emission from an optically thick
wind (\eg\ \citealt{Middleton15}).

For the accretion disk case, the relative normalisations ($N$) of the \diskbb\ and
\diskpbb\ models used for the continuum below 10\,\kev\ can in principle
provide information on the emitting area of these components, as $N \propto
R_{\rm{in}}^{2}/f_{\rm{col}}^{4}$ (where $R_{\rm{in}}$ is the inner radius of the
disk and $f_{\rm{col}} = T_{\rm{col}}/T_{\rm{eff}}$ is the color correction factor
relating the observed `color' temperature to the effective blackbody temperature
at the midplane of the disk, accounting for \eg\ the effects of scattering in the
disk atmosphere on the latter). With the powerlaw (\simpl) model for the
high-energy emission, we find $N_{\rm{cool}} = 110^{+80}_{-60}$ (for \diskbb)
and $N_{\rm{hot}} = 1.0^{+1.0}_{-0.5} \times 10^{-2}$ (for \diskpbb), implying
relative emitting areas of $A_{\rm{hot}}/A_{\rm{cool}} \sim 10^{-4}$ (or
alternatively relative sizes of $R_{\rm{hot}}/R_{\rm{cool}} \sim 10^{-2}$),
assuming the same atmospheric correction for both.

This is very similar to the results obtained by \cite{Miller14} for NGC\,1313 X-1,
and implies a much smaller emitting area for the hotter component. In the case
of the \diskpbb\ component, the statistical uncertainty on the normalization is
driven by a mild degeneracy with the radial temperature index (higher values of
$p$ give larger normalisations). However, for the thin disk scenario ($p=0.75$)
the normalisation is only larger by a factor of $\sim$2, so this degeneracy will
not significantly influence the general conclusion regarding the relative emitting
areas. The relative atmospheric corrections for the two components is probably
a more important issue in this regard. \cite{Shimura95} find $f_{\rm{col}}$ =
1.7 for the disk dominated high/soft state, but it is likely to vary outside of this
accretion regime (see \citealt{Salvesen13, MReynolds13}). Although we do not
know $f_{\rm{col}}$, and it is entirely possible (if not likely) that it would be
different for the hotter and cooler blackbody components, meaning the
quantitative value of their relative emitting areas remains highly uncertain, it
would take an extreme difference in $f_{\rm{col}}$ (\ie\ a factor of $\sim$10)
to reverse the conclusion about their relative areas. 

For the wind model, deviations from the disk geometry assumed in the simple
models used here would result in some additional correction for the area
inferred, and one cannot formally define $f_{\rm{col}}$ in the same manner, as
the wind would not have a mid-plane temperature. However, conceptually
similar atmospheric corrections would still need to be accounted for, and these
differences to the accretion disk case discussed above are also unlikely to be
large enough to reverse the conclusion that the hotter component has a smaller
emitting area than the cooler component.

This is consistent with both the patchy disk and disk/wind models. In the
former, the hotter blackbody component arises from small patches that are
hotter than their surroundings within an inhomogeneous disk, and in the latter
the hotter component arises from the inner disk and the cooler component
arises from a disk wind launched at a larger radius. Further broadband
observations that robustly constrain the relative evolution of the different
emission components are required to differentiate between these interpretations.
We may even find that both patchy accretion discs and large-scale disk winds
contribute significantly to the osberved spectrum, as the presence of one does
not necessarily exclude the other.

Finally, our 2013 campaign cannot constrain the metallicity of the absorbing
medium towards \hoii\ (see section \ref{sec_hardex}). We note that if the
metallicity of this absorbing material is close to solar, as suggested by the work
of \cite{Goad06} and \cite{Winter07}, it is significantly in excess of the
metallicity inferred for the Holmberg II galaxy as a whole by \cite{Egorov13}. If
confirmed, this would support a scenario in which the explosive event in which
\hoii\ was formed enriched its local environment, resulting in a significantly
enhanced local metallicity. Observations with the microcalorimeter due to fly on
\astroh\ (\citealt{ASTROH_tmp}) should have sufficient sensitivity and spectral
resolution to accurately constrain both the oxygen and iron edges associated
with the neutral absorber, and robustly address the metallicity of the absorbing
medium.

\section*{ACKNOWLEDGEMENTS}

The authors would like to thank the referee for their positive feedback, which
helped improve the clarity of the final manuscript. MB and DB acknowledge
financial support from the French Space Agency (CNES). This research has
made use of data obtained with \nustar, a project led by Caltech, funded by
NASA and managed by NASA/JPL, and has utilized the \nustardas\ software 
package, jointly developed by the ASDC (Italy) and Caltech (USA). This
research has also made use of data obtained with \xmm, an ESA science
mission with instruments and contributions directly funded by ESA Member
States, and with \suzaku, a collaborative mission between the space
agencies of Japan (JAXA) and the USA (NASA).

{\it Facilites:} \facility{NuSTAR}, \facility{XMM}, \facility{Suzaku}

\bibliographystyle{/Users/dwalton/papers/mnras}

\bibliography{/Users/dwalton/papers/references}

\label{lastpage}

\end{document}